\documentclass[twocolumn,amsmath,amssymb]{snp}
\usepackage{graphicx}
\usepackage{dcolumn}
\usepackage{bm}
\usepackage{amsmath}
\usepackage{ulem}
\def\nn {\nonumber}

\newcommand{\be}{\begin{equation}}
\newcommand{\ee}{\end{equation}}
\newcommand{\bea}{\begin{eqnarray}}
\newcommand{\eea}{\end{eqnarray}}

\begin{document}
\title{\Large Van der Waals type PV diagrams of PNJL matter}
\author{Manjeet Seth$^1$}
\email{manjeetseth28@gmail.com}
\author{Kinkar Saha$^2$}
\email{saha.k.09@gmail.com}
\author{Sudipa Upadhaya$^3$}
\author{Soumitra Maity$^4$}
\author{Sabyasachi Ghosh$^1$}
\affiliation{$^1$Indian Institute of Technology Bhilai, GEC Campus, Sejbahar, Raipur - 492015, Chhattisgarh, India}
\affiliation{$^2$ Uluberia College, Howrah, West Bengal, India, 711315}
\affiliation{$^3$ Ramsaday College, Howrah, West Bengal, India, 711401}
\affiliation{$^4$ Bose Institute, Block-EN, Sector-V, Salt Lake City, Kolkata, West Bengal, India, 700091}
\maketitle

Study of the systems produced in high-
energy heavy ion collision experiments are of
immense interest for various reasons. Here
we intend to take an attempt to discuss the
presence of interactions in such systems and
their effects to distinguish them from the ideal
scenario under the framework of Polyakov–
Nambu-Jona-Lasinio model. This model en-
capsulates two very important aspects of
Quantum Chromo Dynamics(QCD), viz. the
chiral and the confienement-deconfinement
phase transitions.
With the help of $T$ and $\mu$ dependent constituent quark mass $M (T, \mu)$, one can calculate net quark number density $n(T, \mu)$, pressure $P(T, \mu)$:
\bea
n(T, \mu) &=&g\int \frac{d^3p}{(2\pi)^3}(f^+_0-f^-_0)
\nn\\
P(T, \mu) &=&g\int \frac{d^3p}{(2\pi)^3}\Big(\frac{p^2}{3E}\Big)(f^+_0+f^-_0)~,
\label{nP_T}
\eea
where $g=3\times 3\times 2=18$ is degeneracy factor of quark or anti-quark with energy $E=\sqrt{p^2+M^2(T,\mu)}$ and $f^{\pm}_0=1/\Big\{e^{\beta(E\mp\mu)}+1\Big\}$ for NJL model case and for PNJL model, we have to replaced by modified distribution function~\cite{BI_PNJL1}.

\begin{figure}
  \centering
  \includegraphics[scale=0.24]{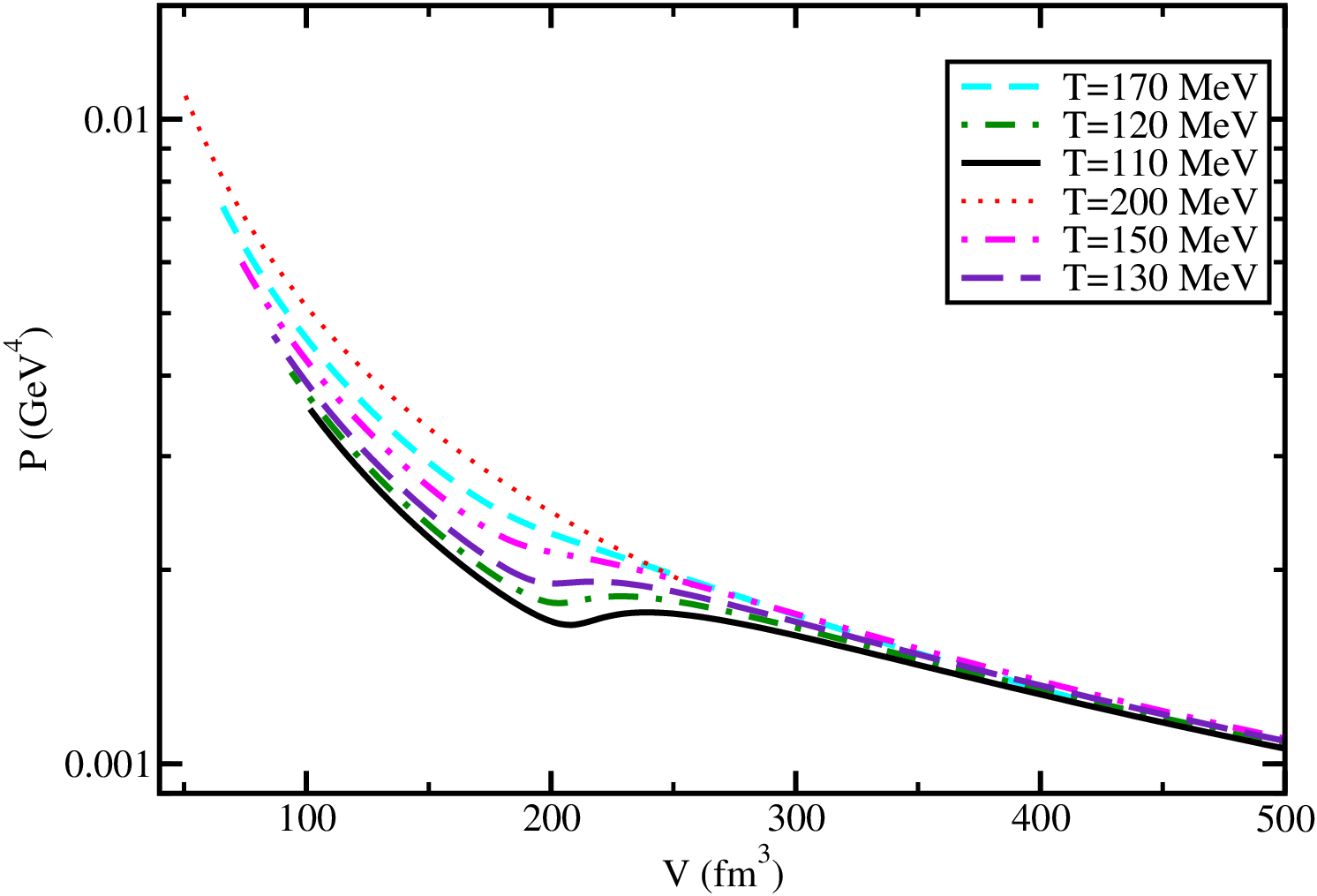} 
  \includegraphics[scale=0.24]{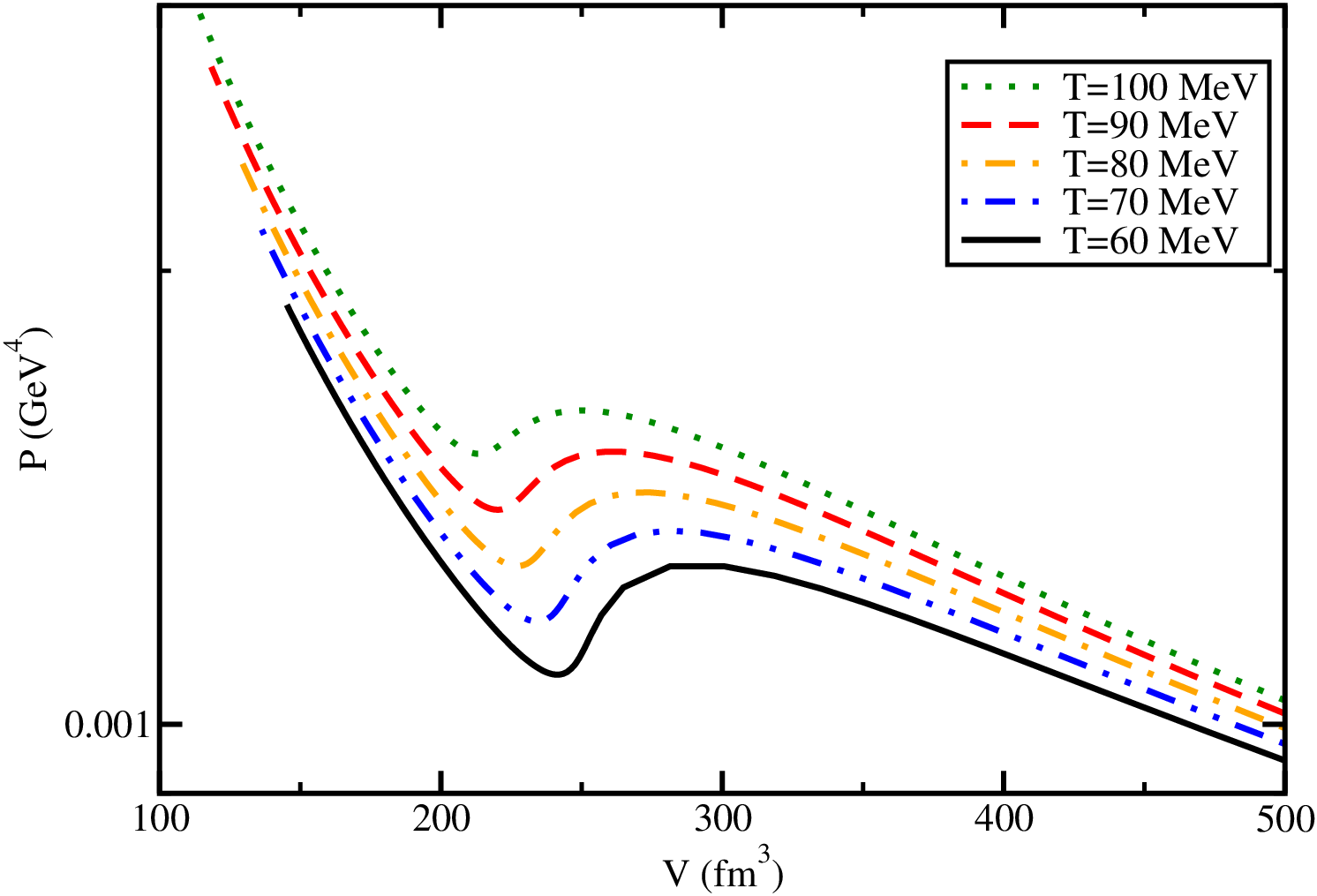}
  \caption{Pressure as a function of volume at various characteristic temperatures by using PNJL model.}
  \label{Andrews}
\end{figure}
Now let us go for rough estimation number of hadrons or quarks, which can be produced in heavy ion collision experiments with large range of $\sqrt{s}=1-200$ GeV. Considering only pion production from this energy, one can get rough no $\frac{(1-200)}{0.14}\approx 7-1,428$ pions and $14-2,856\approx 10^{1-3}$. So if we want to restrict 100 net quark by varrying volume from $V=1$ fm$^3$ to $100$ fm$^3$ with number density $n=100-1$/fm$^3$. Now for 3 temperatures $T=0.400$ GeV, $0.200$ GeV, $0.120$ GeV, fixing net quark number as 
\bea
N=n(T,\mu)\times V&=&100
\nn\\
Vg\int \frac{d^3p}{(2\pi)^3}(f^+_0-f^-_0)&=&100~,
\label{V_mu}
\eea

Fig.(\ref{Andrews}) shows the dependence of Pressure, P on the volume, V of the system, where V is
obtained from Eqn.(\ref{V_mu}) as function of chemical potential, $\mu$.
In order to obtain the constituent masses from PNJL model~\cite{BI_PNJL3,BI_PNJL4,BI_PNJL5} and thereby the volume, V from
Eqn.(\ref{V_mu}) at various characteristic temperatures, we have varied $\mu$. The Pressure, P in Fig.(\ref{Andrews}) show 
non-monotonic behaviour with V. The results are portrayers of the quantum level interactions 
present in the system among the associated degrees of freedom. Also the modified potential
in Polyakov loop part, mimicking to some extent the interactions present among gluonic 
degrees of freedom in Lattice QCD~\cite{BI_PNJL2}, also has a considerable effect [comparison with NJL]. 
From any one isotherm below T=160 MeV, we can get the idea of the corresponding 
$\mu$ where a transition happens indicating a critical point on the phase diagram, using
the V-$\mu$ relation. The transition point is signified by the minima present in the 
corresponding isotherms. For T$>$160 MeV, any such minima are absent as the system then
resides entirely in the QGP region no matter what the chemical potential is. This result
is quite commensurate with the QCD phase diagram and hence T$\sim$160MeV can be considered as the
cross-over transition temperature at $\mu=0$, as extracted from the model study. This
agrees quite satisfactorily with Lattice QCD results at continuum limit~\cite{LQCD1,LQCD2} and as predicted 
by our earlier study~\cite{BI_PNJL2,BI_PNJL3,BI_PNJL4,BI_PNJL5}. 
\par
\begin{figure}
  \centering
  \includegraphics[scale=0.24]{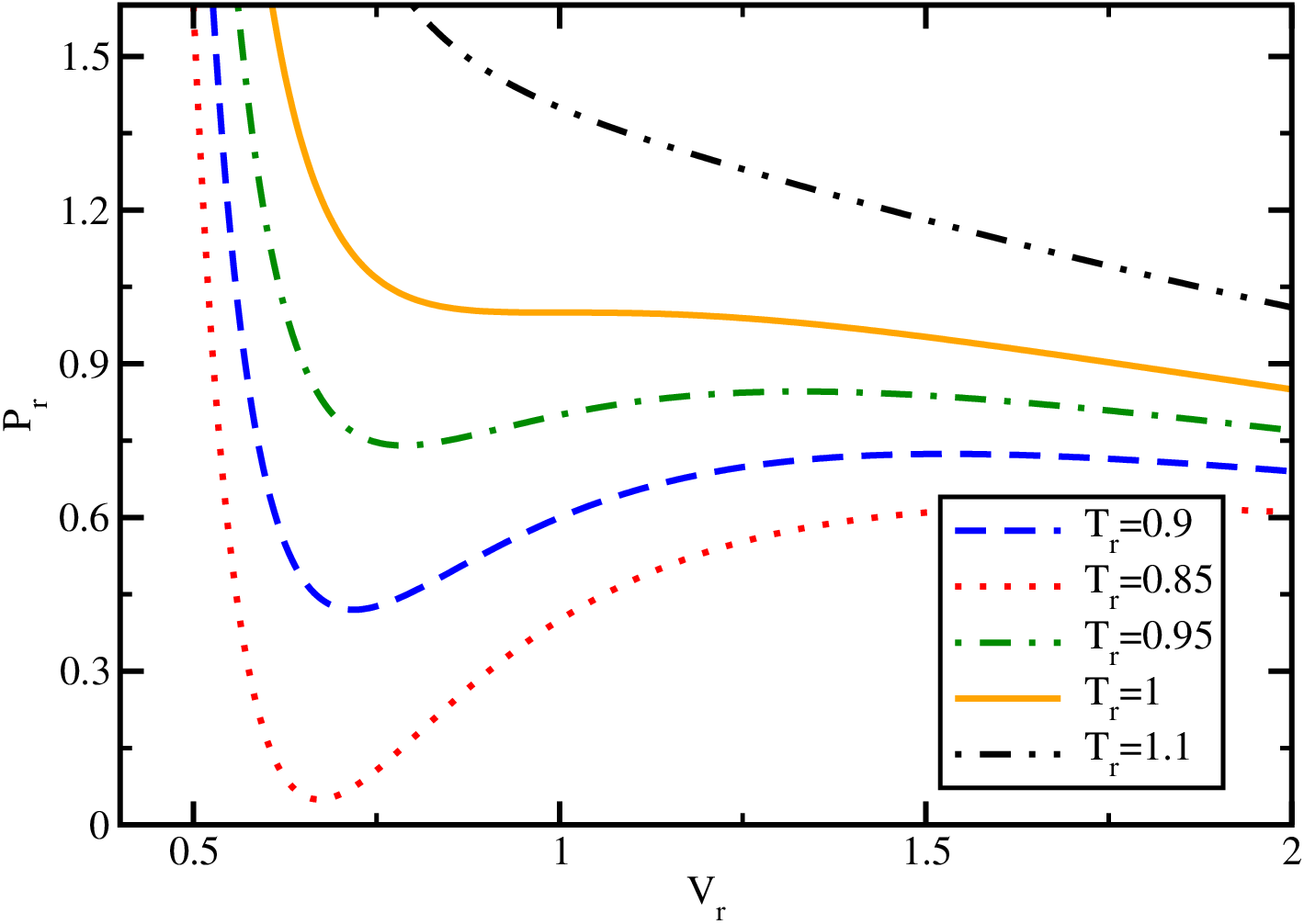} 
  \caption{Van Der Waal like Plots}
  \label{VanDerWaal}
  \end{figure}

The results appear simpler to explain if we look for isotherms with reduced Pressure, $P_r$
and reduced Volume, $V_r$ in Fig.(\ref{VanDerWaal}) as was done by Van Der Waal.
The presence of non-monotonicities below $T_r$=1 signifies the interactions present causing 
transitions occuring in the system at particular chemical potential for one characteristic 
temperature. These critical temperatures and chemical potentials fall on the phase diagram
shown in Ref.~\cite{BI_PNJL2} for six-quark interactions. Below $T_r$=1, the isotherms therefore
represent various states which could be unstable or metastable. Above $T_r$=1, the system 
comprises of quarks only and hence possibility of phase transition does not occur. Similar
results had also been shown in Ref.~\cite{NM}, for normal nuclear matter with corrections from 
Quantum Statistical effects.

\end{document}